\documentclass[aps,pra,twocolumn,unsortedaddress]{revtex4-1}
\usepackage{epsfig,pstricks,graphicx}
\usepackage{amsmath,amssymb}
\usepackage[mathscr]{eucal}
\usepackage{color}
\usepackage[normalem]{ulem}

\def\DJo{$\;$\kern-.4em \hbox{D\kern-.8em\raise.15ex\hbox{--}\kern.35em okovi\'c}}
\def\CC{{\rm\kern.24em \vrule width.04em height1.46ex depth-.07ex
\kern-.30em C}}
\def\RR{{\rm
         \vrule width.04em height1.58ex depth-.0ex
         \kern-.04em R}}
\def\P{{\rm I\kern-.25em P}}
\def\id{{\rm 1\kern-.22em l}}

\newcommand{\beq}{\begin{equation}}
\newcommand{\beqa}{\begin{eqnarray}}
\newcommand{\nbeqa}{\begin{eqnarray*}}
\newcommand{\eeq}{\end{equation}}
\newcommand{\eeqa}{\end{eqnarray}}
\newcommand{\neeqa}{\end{eqnarray*}}
\newcommand{\bra}[1]{\left\langle #1 \right |}
\newcommand{\ket}[1]{\left | #1 \right\rangle}
\newcommand{\braket}[2]{\left\langle #1 | #2 \right\rangle}

\newcommand{\diag}{{\rm diag}\;}

\newenvironment{eqblock}[2]{\beq\label{#2}\begin{array}{#1}}{\end{array}
                                \eeq}
\newenvironment{neqblock}[1]{\[\begin{array}{#1}}{\end{array}\]}
\newcommand{\beqb}{\begin{eqblock}}
\newcommand{\eeqb}{\end{eqblock}}
\newcommand{\nbeqb}{\begin{neqblock}}
\newcommand{\neeqb}{\end{neqblock}}
\begin{document}

\title{Multipartite-entanglement monotones
       and polynomial invariants}
\author{Christopher Eltschka}
\affiliation{Institut f\"ur Theoretische Physik,
         Universit\"at Regensburg, D-93040 Regensburg, Germany}
\author{Thierry Bastin}
\affiliation{Institut de Physique Nucl\'eaire, atomique et de Spectroscopie,
             Universit\'e de Li\`ege, 4000 Li\`ege, Belgium}
\author{Andreas Osterloh}
\affiliation{Fakult{\"a}t f{\"u}r Physik, Campus Duisburg,
Universit{\"at} Duisburg-Essen, 47048 Duisburg, Germany
             }
\author{Jens Siewert}
\affiliation{Departamento de Qu\'{\i}mica F\'{\i}sica, 
             Universidad del Pa\'{\i}s Vasco UPV/EHU,
             48080 Bilbao, Spain}
\affiliation{IKERBASQUE, Basque Foundation for Science, 48011 Bilbao, Spain}

\begin{abstract}
We show that a positive homogeneous function that is invariant under determinant 1 
stochastic local operations and classical communication (SLOCC) transformations
defines an $N$-qubit entanglement monotone if and only if the homogeneous degree is 
not larger than four. 
We then describe a common basis and formalism for the $N$-tangle and 
other known invariant polynomials of degree four. 
This allows us to elucidate the relation of the four-qubit invariants defined by 
Luque and Thibon [Phys. Rev. A {\bf 67}, 042303 (2003)] 
and the reduced two-qubit density matrices of the
states under consideration, thus giving a physical interpretation
for those invariants.
We demonstrate that this is a special case of a completely general law
that holds for {\em any} multipartite system with bipartitions of equal
dimension, {\em e.g.}, for an even number of qudits.
\end{abstract}

\maketitle

{\em Introduction -- }
In recent years an increasing importance of polynomial invariants in the
description of multipartite entanglement has become evident.
It was appreciated in retrospect that both the concurrence~\cite{Wootters1997}
and the
three-tangle~\cite{CKW2000} are polynomial invariants. Originally, the
success of concurrence and three-tangle was based on the lucidity of their
physical concept and the simplicity of their
evaluation, in the case of the concurrence even for arbitrary two-qubit mixed
states~\cite{Wootters1998}. D\"ur {\em et al.}~\cite{Duer2000} proved
that the three-tangle is an {\em entanglement monotone}~\cite{Vedraletal1997,Vidal2000}.
That is, it is a function of the coefficients of a multipartite quantum state which does not
increase {\em on average} under arbitrary
stochastic local operations and
classical communication
(SLOCC) between the parties of a composite quantum system.

Mathematically,  invertible 
local operations on the $j$th subsystem of an $N$-partite quantum system
with local dimensions $d_1,\ldots,d_N$ are represented by the elements of the
group GL$(d_j,\CC)$~\cite{Duer2000,Bennett2000}. While some authors related
concurrence and three-tangle
to hyperdeterminants~\cite{Miyake2002,Briand2003},
the relevance of determinant 1 SLOCC operations had not been realized and exploited until two
seminal papers by Verstraete {\em et al.} appeared~\cite{Verstraete2002,Verstraete2003}.
In Ref.~\cite{Verstraete2003} it was shown that any positive function which is
both invariant under
determinant 1 SLOCC operations and of homogeneous
degree 2 in the wave function coefficients of a pure multipartite quantum state, is necessarily
an entanglement monotone.
At about the same time, 
Klyachko~\cite{Klyachko2002} put forward the interesting idea
to link $N$-qubit entanglement with the notion of semistability of quantum states,
that is the property
that the state can be separated from 0 by a polynomial SL$(2,\CC)^{\otimes N}$ invariant
of its coefficients.

Important mathematical aspects of polynomial
invariants and their relation with multipartite entanglement were investigated, 
{\em e.g.},
in Refs.~\cite{Brylinski2002,Luque2003,Leifer2004,OS2005,Luque2006,Dokovic2009,Li2009,Sharma2010}.
Recently, there is a renewed interest as remarkable new properties of polynomial
invariants have been found such as a new monogamy relation involving the
4-concurrence~\cite{Gour2010a} and SLOCC classifications based on
polynomials~\cite{Viehmann2010}.

We emphasize that for odd qubit number $N$ the lowest degree for a polynomial invariant
is 4, such as in the case of the three-tangle. According to Ref.~\cite{Viehmann2010}
SLOCC classifications may be based on polynomial invariants, in particular on the simple
polynomials of degree 2 and 4. Therefore,
we expect that much more attention
will be devoted to entanglement quantifiers based on such polynomials in the near future.

In this article, we show that a positive homogeneous function invariant under determinant 1 
SLOCC operations defines an  $N$-qubit entanglement monotone 
if and only if the homogeneous
degree is not larger than 4. We recall
known degree-4 polynomials
defined before and demonstrate the relations between them,
thus giving to them a common basis and formalism.
Most interestingly, we can elucidate the relation of  the 
four-qubit invariants of degree 4 defined by
Luque and Thibon~\cite{Luque2003} and the reduced two-qubit density matrices of the
state under consideration.
Finally we show that this is the special case of  an entirely
general statement which holds for any multipartite system with 
bipartitions of equal Hilbert space dimension, such as an
even number of qudits. It comprises also the well-known relation between
concurrence and linear entropy for two qubits~\cite{CKW2000}
and the definition of the $G$-concurrence
for $d\times d$ systems~\cite{Gour2005}.

We start with the extension of an important theorem of Ref.~\cite{Verstraete2003}.

{\em Theorem 1.} We consider a positive homogeneous function $\mu(\psi)$
of the pure multi-qubit state $\ket{\psi}$ that is invariant under local
determinant 1 operations~: $\mu(\lambda \psi) = \lambda^{\eta} \mu(\psi)$ with $\eta, \lambda > 0$.
Then $\mu(\psi)$ is an entanglement monotone if and only if $\eta \le 4$.

{\em Proof.} 
The case $\eta=2$ was proven in Ref.~\cite{Verstraete2003}.
The case $0 <\eta \le 4$ was specifically discussed in Refs.~\cite{Duer2000, Wong2001} 
for the three-tangle and the $N$-tangle, respectively. 
Here we generalize the scope of these proofs to arbitrary invariant homogeneous function 
$\mu$ and we further investigate the case $\eta > 4$.
We consider a two-outcome local positive operator-valued
measure (POVM) on the $k$th party. The two POVM elements $A_1$, $A_2$ obey
$A_1^{\dagger}A_1+A_2^{\dagger}A_2=\id$. They can be written as
$A_j=U_j D_j V$ with unitary matrices $U_j$, $V$ and diagonal matrices
$D_1=\diag{(a,b)}$ and $D_2=\diag{(\sqrt{1-a^2},\sqrt{1-b^2})}$ where $0 \le a,b \le 1$.
For a multipartite state $\ket{\psi}$ the probabilities of the POVM
outcomes are $p_j=\bra{\psi}A_j^{\dagger} A_j \ket{\psi} $. Taking into
account the normalization of the states after application of the POVM,
the homogeneity degree $\eta$  of the considered function $\mu(\psi)$, 
and its invariance under local
unitary operation, $\mu$ is an entanglement monotone if and only if
the inequality
\begin{equation}
     \mu(\psi) \ge p_1 \frac{\mu(D_1 V \psi)}{p_1^{\eta/2}}
                          + p_2 \frac{\mu(D_2 V \psi)}{p_2^{\eta/2}}
\label{monot1}
\end{equation}
is verified for \emph{any} state $|\psi\rangle$ and \emph{any} considered POVM.
We note that $\mu(D_j V \psi)=(\det{D_j})^{\eta/2}\mu(\psi)$ due to the homogeneity 
 and the invariance under local determinant 1 operations.
The normalized state $V \ket{\psi}$ can be written displaying the $k$th qubit
$V \ket{\psi} = \ket{0}_k \ket{\psi_0^{N-1}}+\ket{1}_k \ket{\psi_1^{N-1}}$. Defining
$x \equiv \braket{\psi_0^{N-1}}{\psi_0^{N-1}}$, Eq.~\eqref{monot1} can be rewritten
\begin{equation}
     1 \ge  \frac{(ab)^{\eta/2}}{(xa^2+(1-x)b^2)^{\eta/2-1}}
              + \frac{\sqrt{(1-a^2)(1-b^2)}^{\eta/2}}
                     {(1-xa^2-(1-x)b^2)^{\eta/2-1}}
\label{monot2}
\end{equation}
where $0\le x\le 1$.
We observe that by factoring out $ab$ in the first term in Eq. (2)
and $\sqrt{(1-a^2)(1-b^2)}$ in the second term, the inequality can be
written as
\begin{equation*}
  f_\eta(a,b,x) + f_\eta(\sqrt{1-a^2},\sqrt{1-b^2},x) \leq 1
\end{equation*}
where
\begin{equation*}
  f_\eta(\alpha,\beta,x) = \alpha\beta\left[\frac{\alpha\beta}{x\alpha^2+(1-x)\beta^2}\right]^{\frac{\eta}{2}-1}\ \ .
\end{equation*}
Now for $a,b\neq0,1$, for both terms the base of the exponential in
$f_\eta(\alpha,\beta,x)$ is positive. Since the exponential function for positive
bases is always convex, it follows that
\begin{equation*}
  f_\eta(\alpha,\beta,x) \leq \left(1-\frac{\eta}{4}\right)f_0(\alpha,\beta,x) + \frac{\eta}{4}f_4(\alpha,\beta,x)\ \ .
\end{equation*}
Therefore, if Eq. (2) is true for both $\eta=0$ and $\eta=4$, it holds also for
all values $0<\eta < 4$. For $\eta=0$, a straightforward calculation
shows that the sum in Eq.~(2) gives exactly $1$, and for $\eta=4$, the
inequality was proved by Wong and Christensen in Ref.~\cite{Wong2001},
which concludes our proof for $a,b\neq0,1$.

In order to treat the cases where one of the parameters $a$ or $b$ equals $0$ or $1$, we
note that $f_\eta(\alpha,\beta,x)$ continuously goes to zero if only one of 
$\alpha$ or
$\beta$ goes to zero (and, of course, is also continuous at $\alpha=1$ or $\beta=1$).
Therefore the inequality still holds in this limit. 
Note that this also covers the cases $a=0$, $b=1$ and $a=1$, $b=0$. 
The only remaining cases are 
$a=b=0$ and
$a=b=1$ so that Eq.~\eqref{monot2} is not well defined. But then 
the POVM reduces to a unitary transformation for which the function $\mu$
is constant by definition.
Thus Inequality~\eqref{monot1} is verified for any state and any POVM as long as 
$0 < \eta \le 4$~: $\mu(\psi)$ is an 
entanglement monotone in this case.

Finally we need to show that Eq.~\eqref{monot1} can always be
violated for $\eta>4$. To this end, we consider
an entangled state 
\[
        \ket{\phi}=\alpha
\ket{0}_k \ket{\phi_0^{N-1}}+\beta
\ket{1}_k \ket{\phi_1^{N-1}}
\]
with $\mu(\phi) \neq 0$ where $\ket{\phi_0^{N-1}}$ and $\ket{\phi_1^{N-1}}$ are normalized orthogonal states, and $\alpha>\beta>0$ with
$\alpha^2+\beta^2=1$. We apply a diagonal two-outcome
POVM to $\ket{\phi}$ as in the proof above with $a=\beta/\alpha$ and $b=1$.
By exploiting the relation
$\mu(D_j\phi)=(\det{D_j})^{\eta/2}\mu(\phi)$ we find
for the average value $\bar{\mu}$ after the POVM
\[
\frac{\bar{\mu}}{\mu(\phi)}\ =\
                 2^{-\frac{\eta}{2}+1} \beta^{-\frac{\eta}{2}+2}\alpha^{-\frac{\eta}{2}}\ \ .
\]
It is obvious that for any $\eta>4$ and sufficiently small $\beta$ this ratio can always
be made larger than 1,  thus preventing $\mu$ from being an
entanglement monotone.\hfill $\Box$

Theorem 1 implies in particular that any power of the well-known concurrence 
(or $N$-tangle for $N \ge 3$)
of a state is not an entanglement monotone anymore if it is larger than 2 
(or 1, respectively). 

{\em Various degree-4 invariants. }
In the following we list several known polynomial invariants
of degree 4 and highlight the relations between them.
We write the $N$-qubit state $\ket{\psi}$ in the standard basis
$\ket{\psi}=\sum a_{i_1 \ldots i_N}\ket{i_1 \ldots i_N}$.
In Ref.~\cite{Wong2001}, Wong and Christensen defined
the $N$-tangle 
\begin{eqnarray}
   \tau_{N}&=&2\left|\sum  a_{\alpha_1 \ldots \alpha_N}
                                 a_{\beta_1 \ldots \beta_N}
                                 a_{\gamma_1 \ldots \gamma_N}
                                 a_{\delta_1 \ldots \delta_N}
 \right.
\nonumber\\
                  && \times \epsilon_{\alpha_1\beta_1}
                            \epsilon_{\alpha_2\beta_2}\ldots
                            \epsilon_{\alpha_{N-1}\beta_{N-1}}
                            \epsilon_{\gamma_1\delta_1}
                            \epsilon_{\gamma_2\delta_2}\ldots
\nonumber\\
                  && \left.
                     \times \epsilon_{\gamma_{N-1}\delta_{N-1}}
                            \epsilon_{\alpha_N\gamma_N}
                            \epsilon_{\beta_N\delta_N}\right|
\label{Wong}
\end{eqnarray}
where $\epsilon_{01}=-\epsilon_{10}=1$ and $\epsilon_{00}=
\epsilon_{11}=0$. Note that
the three-tangle~\cite{CKW2000} is obtained for $N=3$.

A method to systematically construct SL(2,$\CC)^{\otimes N}$-invariant
$N$-qubit polynomials was developed in Ref.~\cite{OS2005}.
Now we show that the formalism defined  there  
provides a transparent way to write also the Wong-Christensen invariants $\tau_N$.
With the notation of Ref.~\cite{Dokovic2009} they can
be written as
\begin{equation}
\mathcal B^{(N=2k+1)}_{(1)}=
      ((\sigma_\mu  \sigma_2 \ldots \sigma_2 \bullet \sigma^\mu \sigma_2\ldots  \sigma_2))
\label{B-odd}
\end{equation}
(with $2k$ operators $\sigma_2$ on each side of the $\bullet$ symbol and the lower
index indicating the position of the contraction from 1 to $N$) for odd $N$ and
\begin{equation}
\mathcal B^{(N=2k)}_{(1,2)}=
      ((\sigma_\mu \sigma_\nu \sigma_2 \ldots \sigma_2 \bullet
                              \sigma^\mu \sigma^\nu \sigma_2\ldots  \sigma_2))
\label{B-even}
\end{equation}
(with $2k-2$ operators $\sigma_2$ on either side of $\bullet$) for even $N$, respectively.
Note that for even $N$ two contractions
are necessary. Their positions $(1,j)$ ($1<j\le  N$) are given in the lower indices
of $B^{(N=2k)}_{(1,j)}$.
Here we have used the following definitions:
\begin{align}
(( A_1 \bullet A_2 )) &\ =\
     \left\langle \psi^{\ast} |  A_1  \psi \right\rangle
     \left\langle \psi^{\ast} | A_2 \psi \right\rangle \ \
\\
\sigma_\mu \bullet \sigma^\mu & \ =\
    \sum_{\mu=0}^3{g_\mu \cdot \sigma_\mu \bullet \sigma_\mu}
\label{summation}
\end{align}
for operators $A_i$ that act on the Hilbert space of $\psi$,
the Pauli matrices
$(\sigma_0,\sigma_1,\sigma_2,\sigma_3) = (\id_2,\sigma_x,\sigma_y,\sigma_z)$ and
$(g_0,g_1,g_2,g_3):=(-1,1,0,1)$. The $\bullet$ symbol stands for
a tensor product related to copies of the same state whereas
we do not write explicitly tensor products between the parties:
$ \ldots \sigma_\mu \sigma_\nu \ldots  \equiv
                  \ldots \sigma_\mu \otimes \sigma_\nu \ldots
$
In general, these SL(2,$\CC)^{\otimes N}$-invariant polynomials are {\em not} invariant under
qubit permutations.
One obtains more degree-4 invariants from Eqs.~\eqref{B-odd},\eqref{B-even} by
permutation of the qubits and/or by replacing  $\sigma_2 \bullet \sigma_2$
for a given qubit with $\sigma_{\mu} \bullet \sigma^{\mu}$ 
(see Ref.~\cite{Dokovic2009}).
It is equally well possible to define symmetric polynomials
by means of appropriate symmetrization as proposed, {\em e.g.}, in
Refs.~\cite{Dokovic2009,Li2009-2}.

The $N$-tangle $\tau_N$ turns out to be a special case of the $\mathcal B$ invariants.
To show this we note the important relations
\begin{eqnarray}
\epsilon_{ij}\epsilon_{kl} &=& -  \bra{ik} \sigma_2\otimes \sigma_2 \ket{jl}\  ,
\nonumber\\
\epsilon_{ik}\epsilon_{jl} &=&  -\frac{1}{2} \sum_{\mu} \eta_{\mu}
                                    \bra{ik} \sigma_{\mu}\otimes \sigma_{\mu} \ket{jl}
\nonumber
\end{eqnarray}
with the Minkowski-like metric $\eta_{\mu}=(-1,1,1,1)$
which, after substitution into Eq.~\eqref{Wong}, lead to
\begin{equation}
    \tau_N=\left|\sum_{\mu} \eta_{\mu} \bra{\psi^{\ast}} \sigma_2^{\otimes N-1}\otimes 
                                                                     \sigma_{\mu} \ket{\psi}
                      \bra{\psi^{\ast}} \sigma_2^{\otimes N-1}\otimes\sigma_{\mu} \ket{\psi}\right|
\label{Wongzwischen}
\end{equation}
Hence, for odd $N$ we find immediately
\begin{equation}
 \tau_{N=2k+1}\ \equiv\ \left| {\mathcal B}^{(N)}_{(N)}\right|
\end{equation}
as the term $\mu=2$ term in the sum of Eq.~\eqref{Wongzwischen} vanishes~\cite{OS2005}.
On the other hand, for even $N$ only the $\mu=2$ term in the sum survives and we
recover the well-known
result that the Wong-Christensen tangle equals (up to a prefactor) the square of the
{\em N-concurrence}~\cite{Wong2001,Bullock2004}
\begin{equation}
 \tau_{N=2k}\ \equiv \   \left| (( \sigma_2^{\otimes N}\bullet \sigma_2^{\otimes N}))\right|
                       =  \left|(( \sigma_2^{\otimes N}))\right|^2
\end{equation}
and can be considered the  $\left|{\mathcal B}^{(2k)}_{(0)}\right|$ 
invariant without any contractions.

We mention that the degree-4 invariants for $N$ qubits form a vector space
of dimension $(2^{N-1} +(-1)^N)/3$ (see Ref.~\cite{Brylinski2002}). Due to Theorem 1
the absolute value of any polynomial in this space is an entanglement monotone.

{\em Four qubits. - }
For $N=4$, {\em e.g.}, the polynomials $\mathcal B^{(4)}_{(1,2)}$,
$\mathcal B^{(4)}_{(1,3)}$ and $\mathcal B^{(4)}_{(1,4)}$ may be used as the
basis polynomials.
Alternatively, three four-qubit invariants $L$, $M$, $N$ were introduced 
by Luque and Thibon~\cite{Luque2003}
via the determinant
\begin{equation}
          L\ =\ \left|
                \begin{array}{cccc}
                a_{0000} & a_{0100} & a_{1000} & a_{1100}\\
                a_{0001} & a_{0101} & a_{1001} & a_{1101}\\
                a_{0010} & a_{0110} & a_{1010} & a_{1110}\\
                a_{0011} & a_{0111} & a_{1011} & a_{1111}
                \end{array}
                \right|
\label{LTL}
\end{equation}
and $M$, $N$ analogous with the second and the third, or the second and the fourth
qubit exchanged, respectively. These invariants
are related to ${\mathcal B}^{(4)}_{(1,j)}$ via~\cite{Dokovic2009} 
\begin{eqnarray}
L&=(1/48)({\mathcal B}^{(4)}_{(1,3)}-({\mathcal B}^{(4)}_{(1,4)})\ \ ,
\nonumber\\
M&=(1/48)({\mathcal B}^{(4)}_{(1,4)}-({\mathcal B}^{(4)}_{(1,2)})\ \ ,
         \\
N&=(1/48)({\mathcal B}^{(4)}_{(1,2)}-({\mathcal B}^{(4)}_{(1,3)})\ \ ,
\nonumber
\end{eqnarray}
that is, they are linearly dependent, $L+M+N=0$.

It turns out that the invariants $L$, $M$, $N$ are closely
related to the two-qubit reduced density matrices of the original pure
four-qubit state $\ket{\psi}$:
\begin{equation}
       | L |^2 = \det{[\mathrm{tr}_{34}(\ket{\psi}\!\bra{\psi})]} \equiv \det{\rho_{12}}
       \ \ .
\label{idL1}
\end{equation}
where the $\rho_{12}$ is obtained from $\ket{\psi}$ 
by tracing out the third and the fourth qubit.
For $M$ and $N$ we have the analogous relations
\begin{equation}
       | M |^2 = \det \rho_{13} \ \ \ , \ \ \ \
       | N |^2 = \det \rho_{14}
\label{idL3}
\end{equation}
with $\rho_{13}\equiv \mathrm{tr}_{24}(\ket{\psi}\!\bra{\psi})$
and  $\rho_{14}\equiv \mathrm{tr}_{23}(\ket{\psi}\!\bra{\psi})$.
We proceed by proving Eq.~\eqref{idL1}, the proof for Eq.~\eqref{idL3}
is analogous.
To this end, it is essential to note that the reduced density matrix
$\rho_{12}$ can be written as a matrix product 
$\rho_{12}=X^{\dagger}X$~\cite{Verstraete-vEnk2003}.
This can be seen as follows. We write the pure state as
$\ket{\psi}=\sum_{i,k} a_{i,k} \ket{i,k}$ where the two-digit binary indices
$i$ and $k$ run from $00$ to $11$. The reduced density matrix of the first
two qubits is given by
\begin{equation}
    \rho_{12}=\sum_{i,k}\sum_l  a_{i,l} a^{\ast}_{k,l} \ket{i}\!\bra{k}\ \ .
\end{equation}
Obviously, the coefficients of $\rho_{12}$ are given by a matrix product
$X^{\dagger}X$ with $(X^{\dagger})_{i,l}=a_{i,l}$. The latter matrix is just the transposed
of the matrix in Eq.~\eqref{LTL}. Thus, we have proven the identification in
Eqs.~\eqref{idL1}. 
Consequently, the Luque-Thibon invariants, which up to now
seemed to represent an arbitrary choice of degree-4 invariants, are seen 
to have a direct physical meaning: They carry specific information 
about the entanglement of half of the qubits in a pure four-qubit state
with the remaining ones.

We may add two remarks which lead directly to a generalization of this result.
{\em i)} The two-qubit analog of this statement is the well-known fact
that the squared concurrence of a pure two-qubit state equals the linear
entropy of either qubit in that state~\cite{CKW2000}.
{\em ii)} Our proof provides an alternative confirmation that $L$ is
an SL$(2,\CC)^{\otimes 4}$ invariant for $\ket{\psi}$. Let us consider
the bipartition of first and second, and third and fourth qubit,
respectively. The determinant of $\rho_{12}$ equals the product of
the Schmidt coefficients for this $4\times 4$ state. It is not changed
by SL$(4,\CC)$ operations on the first four-dimensional partition.
On the other hand, this determinant equals the one of $\rho_{34}$
(the reduced density matrix of the third and fourth qubit of $\ket{\psi}$)
which again is an SL$(4,\CC)$ invariant.

{\em Arbitrary $d\times d$ systems. - } The preceding remarks are clearly
not limited to pure states of four-qubits. In fact, they can readily be
extended to arbitrary $d\times d$ systems by noting that all steps in the
proof of Eq.~\eqref{idL1} can be applied, one merely
has to change the range of the indices $i$, $k$ 
into $0,\ldots,(d-1)$. Thus we have  the next theorem~:

{\em Theorem 2.} Given the pure state $\ket{\psi}$
of a composite system with a $d\times d$ bipartition $\ket{\psi}=\sum_{i,k}
a_{i,k}\ket{i,k}$  $(i,k=0,\ldots,d-1)$,  the determinant of
the (`reshaped') coefficient matrix $X^{\dagger}\equiv(a_{i,k})$
always defines a polynomial SL$(d,\CC)$ invariant $\nu(\psi)$ of homogeneous degree $d$:
\[
         \left| \nu(\psi) \right|^2\ = \ \det\rho_{[d]}\ \equiv\ \det{X^{\dagger}X}
\]
where $\rho_{[d]}$ denotes the reduced density matrix of $\ket{\psi}$ obtained by
tracing out one $d$-dimensional bipartition.

For a system with an even number $N$ of qubits, we obtain
$N!/(2 (N/2)!^2)$ degree-$N$ invariants from this theorem.
On the other hand, for a bipartite system of $d\times d$ dimensions
it defines a unique degree-$d$ invariant whose absolute value
with an appropriate exponent $\alpha$ gives an entanglement monotone.
For $d=2$,  $\alpha=1$ (two qubits) it is
identical to Wootters' concurrence while for $d>2$, $\alpha=1/d$ it is the
$G$-concurrence~\cite{Gour2005}.

The theorem cannot easily be extended to
$d\times d^{\prime}$ systems with $d\neq d^{\prime}$. In that case
the determinant for the reduced density matrix of the subsystem with
larger dimension vanishes while that of the lower-dimensional subsystem
in general does not.

{\em Conclusion. - } We have discussed the relations between various
degree-4 polynomial SL(2,$\CC)^{\otimes}N$ invariants of $N$-qubit states. In particular,
we have found that the Wong-Christensen invariants are special cases
of more general degree-4 invariants that can be obtained with the formalism
in Refs.~\cite{OS2005,Dokovic2009}. 
We have shown in Theorem 1 that {\em any} positive homogeneous
SL(2,$\CC)^{\otimes N}$-invariant function with
positive homogeneity degree up to 4  is an entanglement
monotone while it is not for  larger degrees. 
This yields an upper bound to the power of any homogeneous 
SL(2,$\CC)^{\otimes N}$-invariant entanglement monotone 
that can be considered 
without losing the monotonicity property.
This result is satisfactory also as it shows that
for all qubit numbers there exist many polynomial entanglement
monotones (recall that the lowest possible polynomial degree is 4
for odd qubit number). 

We have then elucidated the physical
meaning of the four-qubit invariants of degree 4. We have proven that
the peculiar linear combinations found by Luque and Thibon~\cite{Luque2003}
are related to the two-qubit reduced density matrices of the pure four-qubit
state. 
Thus they provide information about the entanglement of any two 
qubits in the state with the other two. In this way, 
the Luque-Thibon invariants play a role for four qubits which is  analogous to that 
of the concurrence for two qubits.
Most importantly, it was straightforward to extend this finding in Theorem 2
to any system
with a $d\times d$ bipartition, that is, in particular to systems with
even qubit number $N$ and to bipartite $d\times d$ systems.
The striking feature of Theorem 2 is that it links previously unrelated
facts such as the monogamy relation for pure two-qubit states, the
existence of the $G$-concurrence as an entanglement measure,
and the Luque-Thibon invariants for pure four-qubit states.

{\em Acknowledgments. -- }
We thank S.\ Szalay for pointing out an error in a  previous
version of the proof for Theorem 1.
This work was supported by the German Research Foundation within SFB 631
and SPP 1386 (CE), and the Basque Government grant IT-472 (JS).
CE and JS thank J.\ Fabian and  K.\ Richter (University of Regensburg)
for their support of this research.

\end{document}